\documentclass[
    ,final            
  ]
  {aipproc}

\layoutstyle{8x11double}

\usepackage{epsfig}

\newcommand{\sinatm}{\ensuremath{\sin^2 2 \theta_{13}}}
\newcommand{\sinsol}{\ensuremath{\sin^2 2 \theta_{12}}}

\newcommand{\matmabs}{\ensuremath{|\Delta m^2_{31}|}}
\newcommand{\msol}{\ensuremath{\Delta m^2_{21}}}

\newcommand{\sint}{\ensuremath{\sin^2 2 \theta_{13}}}
\newcommand{\dcp}{\ensuremath{\delta_{cp}}}

\begin{document}

\title{Sensitivities of Future Long Baseline Experiments in the U.S.}


\author{Mark Dierckxsens}{
  address={The Enrico Fermi Institute, The University of Chicago, 
    Chicago, Illinois, 60637}
}

\begin{abstract}
  
  Sensitivities to neutrino oscillation parameters for possible very long 
  baseline neutrino oscillation experiments are discussed.
  The reach for observing a non-zero mixing angle $\theta_{13}$, establishing
  CP violation and determining the mass hierarchy are compared between various 
  experimental options. 
  Different possibilities for neutrino beams are briefly described, as well
  as the assumptions about the performance of a large water Cherenkov and 
  liquid Argon detector.

\end{abstract}

\maketitle

\section{Introduction}

Observation of a non-zero value for $\theta_{13}$ opens the door for a 
measurement of CP violation in the lepton sector and the determination of 
the neutrino mass hierarchy.
Several experiments are expected to reveal more information on this angle
in near future reactor \cite{doublechooz,dayabay} and mid term accelerator
\cite{t2k,nova} experiments.
An extensive program is required to study neutrino oscillations beyond the
scope of these experiments.

A joint study between Brookhaven National Lab  (BNL) and Fermi National 
Accelerator Lab (FNAL) was organized to investigate the potential of a very 
long baseline experiment in the United States. 
The conclusions of this elaborate effort are discussed in 
\cite{Barger:2007yw}, wherein references to supporting documentation 
can be found (see also \cite{reporturl}).%

Most calculations were performed using the GLoBES \cite{globes}
software package. 
The basis for the inputs will be described in the next sections, after
which the results will be discussed.

\section{Neutrino Beams and Baselines}

One possibility is to use the existing NuMI beamline and place a 
detector at 810km about $0.8^{\circ}$ off-axis.
Simulations of the expected spectra are base upon GNuMI which is
well tested against experimental data \cite{minos}.

The same Monte Carlo is used simulate the Wide Band Low Energy (WBLE) beam 
with a few modifications: a new design for the horns and a 400m long, 4m 
diameter decay pipe. 
Incident proton energies of 120 GeV were preferred over lower energy
beams due to a higher total neutrino yield. 
The beam has been tilted slightly off-axis by $0.5^{\circ}$ to maintain a 
large fraction of the low energy flux but to suppress the high energy tail.
Results quoted use this configuration, although calculations were
also performed for other options in this study.

The default assumes a total of $6 \times 10^{21}$ 
protons-on-target (PoT) are delivered to produce above mentioned beams,
divided equally among the neutrino and anti-neutrino running mode.

Various baseline lengths were considered for the WBLE beam but only 
results for 1300km will be shown.
This is the distance from FNAL to the Homestake mine, which 
recently was singled out as the candidate for a Deep Underground Science 
and Engineering Laboratory.

\section{Neutrino Detectors}

A comparison is made between two detector technologies: the well 
established water Cherenkov (WCh) technique on the one hand, and a 
Liquid Argon (LAr) detector on the other hand.
The parametrization of the detector responses used as input 
for the sensitivity calculations are described in this section.

\subsection{Water Cherenkov Detector}

The detector performance is based on the elaborate work performed to reduce
the contamination from NC interactions in the $\nu_e$ 
selection procedure using SuperK Monte Carlo events \cite{yanagisawa}. 
Quantities obtained from a significantly improved $\pi^0$ 
finding algorithm are combined with other variables in a 
likelihood based selection method. 
This increases the signal-to-background ratio
drastically in addition to the standard SuperK cuts to select single ring 
electron-like events inside the fiducial volume.
The efficiencies used for the calculations reported here are matched to
these results.
An independent group confirmed these findings using similar techniques
\cite{dufour}.

The energy smearing function for $\nu_e$ signal events 
is based on a GEANT3 \cite{geant} simulation of quasi-elastic interactions 
and is about $10\%/\sqrt{E}$ at 1~GeV. 
It is in very good agreement with the results from the two studies 
mentioned above, except for missing asymmetric tails due to 
non-quasi elastic events.
However, this is shown to have a small effect on the 
final result \cite{dufour}.
The resolution functions for NC interactions are based on the 
NUANCE Monte Carlo program \cite{nuance}.

\begin{figure}[tbp!]
  \centering
  \epsfig{file=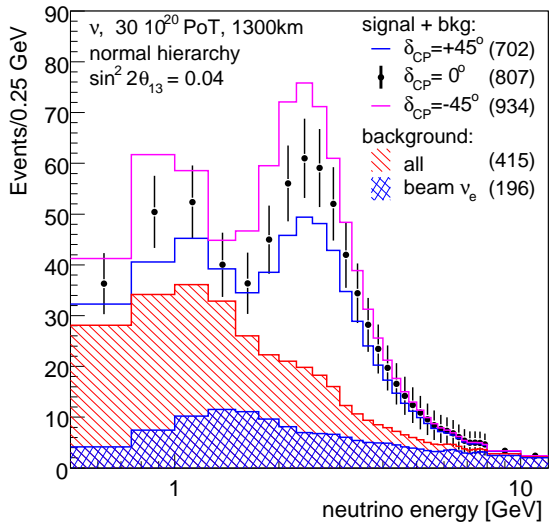,width=0.45\textwidth}
  \caption{The expected $\nu_e$ appearance spectrum for a 300~kton WCh 
    detector 
    for $\sint=0.04$ exposed to $3 \times 10^{21}$ PoT  
    producing a neutrino beam. The points assume $\dcp=0^{\circ}$ and 
    normal mass hierarchy, while the solid lines have different $\dcp$ 
    values. The different background contaminations are indicated by 
    the hatched histograms.
    Total event rates for each variation are given between brackets. 
}
  \label{fig1}
\end{figure}

Figure~\ref{fig1} shows the expected spectrum using a 300~kton fiducial volume
water Cherenkov detector.

\subsection{Liquid Argon Detector}

A Liquid Argon detector is often considered as the best possible neutrino 
detector due to a fully active calorimetric medium with fine grained 
tracking capabilities. 
This should allow for a high signal selection efficiency while 
practically eliminating all neutral current backgrounds.

Based on a hand scanning study  \cite{lartpc}, the selection efficiency 
for charged current $\nu_e$ interactions is taken as $80\%$ with a complete 
rejection of NC background events. 
The energy resolution for quasi-elastic events is assumed to be 
$5\%/\sqrt{E}$, while this is $20\%/\sqrt{E}$ for all other type of 
interactions.

The fiducial mass of the LAr detector under consideration is 100 kton.

\section{Sensitivity Calculations}

Three limits as a function of true \sint\ and \dcp\ are derived for each 
experimental variation: sensitivity to a non-zero $\sint$, CP violation and 
resolving the mass hierarchy. 
When single value limits are quoted, this corresponds to the value of $\sint$\
above which the  sensitivity is better than the quoted confidence level
for $50\%$ of the \dcp\ phase space. 
For a limited set of $(\sint,\dcp)$ combinations, the ability to measure
these values is also explored.
Correlations between oscillation parameters are taken into account.

The values of the atmospheric mixing parameters were chosen to be
$\matmabs = 2.7 \times 10^{-3} eV^2$ and $\sinatm = 1$, 
with the error determined from the $\nu_{\mu}$ disappearance channel. 
The solar parameters are set to $\msol = 8.6 \times 10^{-5}$ and 
$\sinsol = 0.86$, both with an error of $5\%$.
The error on the matter density is taken to be $5\%$, while the 
error on the background is assumed to be $10\%$.

The obtained limits for each experiment described above are listed in 
table~\ref{tab:sens}. 
The overall best performance is obtained with a LAr detector placed in
the WBLE beam. 
When placed in the NuMI off-axis beam, the sensitivity to CP violation is
still good, but the ability to determine the mass hierarchy is much worse.
The WCh detector performs good for all three hypotheses tested.

\begin{table}[tbp]
  \centering
  \begin{tabular}{c|ccc}
    & $\sint \neq 0$ & CP violation  & mass hierarchy \\
    \hline
   WCh-W& 0.006 & 0.03  & 0.01  \\
   LAr-W & 0.003 & 0.005 & 0.006 \\
   LAr-N & 0.002 & 0.03  & 0.05  \\
  \end{tabular}
  \caption{Sensitivities to a non-zero \sint, CP violation and the mass 
    hierarchy at the $3\sigma$ level for a WCh detector in the WBLE beam 
    (WCh-W), a LAr detector in the WBLE (LAr-W) and NuMI off-axis beam 
    (LAr-N).
    Limits are obtained after a total exposure of $6 \times 10^{21}$ 
    PoT equally divided among $\nu$\ and $\bar{\nu}$\ running.}
  \label{tab:sens}
\end{table}

\begin{figure}[tbp]
  \centering
  \epsfig{file=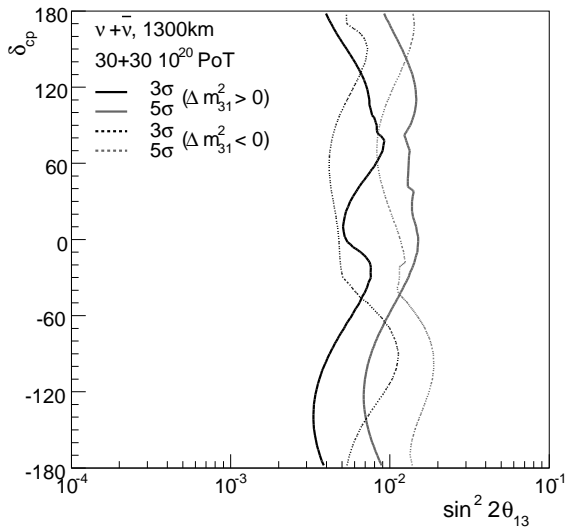,width=0.45\textwidth}
  \caption{The sensitivity to the mass ordering for a LAr detector exposed 
    to a total of $6 \cdot 10^{21}$ PoT producing the WBLE beam.
    For true values of \sint\ and \dcp\ to the right of the solid (dotted) 
    lines, the inverted (normal) mass hierarchy can be excluded. The black 
    (gray) lines show the sensitivity at the $3\sigma$ ($5\sigma$) level.}
  \label{fig2}
\end{figure}

The sensitivity to the mass hierarchy as function of true \sint\ and
\dcp\ is shown in figure~\ref{fig2} for the LAr detector.
The dependence on \dcp, which is already small, can be further reduced 
by running longer in anti-neutrino mode. 

\begin{figure}[tbp]
  \centering
  \epsfig{file=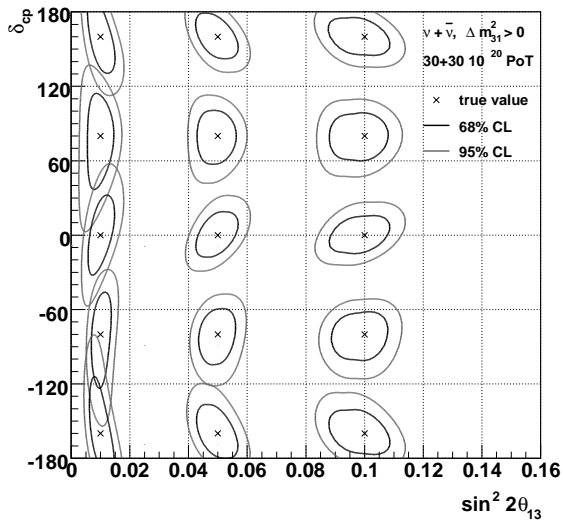,width=0.45\textwidth}
  \caption{The measurement of \sint\ and \dcp\ for a WCh detector 
    assuming an exposure to $6 \times 10^{21}$ PoT.
    The true input values are indicated by the crosses, while the contours 
    are shown at $68\%$ (black) and $95\%$ (gray) confidence level.}
  \label{fig3}
\end{figure}

As shown in figure~\ref{fig3}, \sint\ can be measured to $10\%$ at $68\%$ 
C.L.\ with a WCh detector for values larger than $0.01$ 
independently of \dcp\ in the case of normal hierarchy. 
For a LAr detector in the WBLE beam, this number decreases to $6\%$.
The errors are very similar if the true mass hierarchy turns out
to be inverted.

\subsection{Background Uncertainties and Exposure}

The sensitivities for a WCh detector excluding systematic errors are listed
in table~\ref{tab:sens-2}.
It is clear that the CP exclusion limit is most affected by the background 
uncertainty.
This measurement can benefit significantly if this uncertainty can be 
reduced to $5\%$.
This seems perfectly reasonable as it is almost certain  a near detector 
will be deployed which will measure the background contamination to high 
precision.
Table~\ref{tab:sens-2} also shows the results when the exposure is doubled.

\begin{table}[tbp]
  \centering
  \begin{tabular}{c|ccc}
    $\quad \quad$ & $\sint \neq 0$ & CP violation  & mass hierarchy \\
    \hline
    I & 0.005 & 0.01  & 0.01  \\
    II  & 0.005 & 0.02 & 0.01 \\
    III & 0.004 & 0.01 & 0.008
  \end{tabular}
  \caption{Sensitivities for a WCh detector under different 
    assumptions: 
    I) statistics only, II)  a $5\%$ uncertainty on the background and 
    III) same as II but for twice the running time, i.e.\ $6 \times 10^{21}$ 
    PoT for neutrino and anti-neutrino running each.}
  \label{tab:sens-2}
\end{table}

\section{Conclusions}

Details of the sensitivity calculations performed in the scope of the 
joint FNAL-BNL U.S.\ long baseline neutrino experiment study were 
discussed. 
The best limits for the same exposure are obtained using a 100~kton
LAr detector in the WBLE beam. 
Placing this detector off-axis in the NuMI beam reduces the sensitivity
to the mass hierarchy significantly.
The 300~kton WCh  detector placed at the DUSEL candidate site has a 
good overall performance.
It was also shown that this detector is not limited
by background systematics if controlled to better than  $5\%$.

\begin{theacknowledgments}
  I am grateful to Brookhaven National Laboratory for supporting the 
  work presented. 
  I want to thank everybody involved in this study for their valuable 
  input, and in particular Milind Diwan and Mary Bishai for their 
  fruitful discussions.
\end{theacknowledgments}

\end{document}